\documentstyle[12pt]{article}

\newcommand{\be}{\begin{equation}}
\newcommand{\ee}{\end{equation}}

\newcommand{\beba}{\begin{equation}\begin{array}{lll}}
\newcommand{\eaee}{\end{array}\end{equation}}
\newcommand{\bea}{\begin{eqnarray}}
\newcommand{\eea}{\end{eqnarray}}
\newcommand{\ba}{\begin{array}}
\newcommand{\ea}{\end{array}}

\newcommand{\norsl}{\normalsize\sl}
\newcommand{\norsc}{\normalsize\sc}
\begin{document}
\textheight 22cm
\voffset  -1cm
\begin{titlepage}
\title{\Large\bf { Finite Temperature Spacelike Gluon Propagators
in Lattice Momentum Space }}
\author{                               
\norsc G. Koutsoumbas  \\
\\
\norsl  Physics Department\\
\norsl  National Technical University\\
\norsl  157 80 Zografou, Athens, Greece\\}

\date{}
\maketitle

\begin{abstract}
{\normalsize \noindent We study the behaviour of lattice momentum-space gluon
propagators for a pure $SU(2)$ gauge theory at finite temperature. We find
out that the magnetic mass is $0.26 g^2(T) T$; we have repeated the
same calculations in three dimensions.

}
\end{abstract}

\begin{picture}(100,100)(0,-287)                                       
\put(0,-300){NTUA 49/96}
\put(0,-287){January 1996}
\end{picture}

\thispagestyle{empty}
\end{titlepage}

\section{Introduction}
\noindent

The subject of the finite temperature behaviour of the gluon propagators
has attracted attention due to its relevance in understanding processes in
the early universe. Very early approaches \cite{gross,linde} have
identified the two different kinds of masses arising when considering
longitudinal or transverse degrees of freedom, called electric and magnetic
masses respectively.
It has been argued \cite{rebhan} that they are gauge
invariant, thus measurable quantities.

The pole for the electric gluon propagator may be obtained in
perturbation theory, with the well-known one-loop result:
\be
m_e =  \sqrt{\frac {2 N_c+N_f}{6} } g(T) T, \label{elmass}
\ee
for $N_f$ massless quark flavours and the $SU(N_c)$ gauge group.
This has been measured numerically using the heavy quark potential and
the correlations of Polyakov lines \cite{nadkarni,bac}.

The magnetic gluon propagator has no (non-zero) pole at the one-loop
level, however at two loops there is a contribution to the magnetic
mass proportional to $g^2(T) T$. Higher loops contribute multiples of
the same quantity, leaving no possibility for perturbative
calculations. A non-zero value for this non-perturbative quantity may
soften the severe infrared problems of finite temperature perturbative
calculations, since it will act as an infrared cut-off. The subject of the
magnetic mass has been attacked since the eighties, using various methods.
On the lattice, it has been measured using the effects of twisted boundary
conditions on bulk quantities \cite{bls,degrand} and yielded the value
$0.24 g^2(T) T$. The next approach \cite{mandula} has been more
straightforward, in the sense that it studied the correlators of
(gauge-variant) gluon operators. It found effective masses increasing with
distance, being thus in conflict with the K\"allen-Lehmann representation.
The interpretation was that this odd behaviour can be acceptable for a
confined state (such as the gluon), although it should be rejected for a
physical particle.

Restricting ourselves to pure gauge theories, we have
three scales that enter the game: the temperature T, the scale of the 
electric mass, $g(T) T$, and the scale of the magnetic mass, $g^2(T) T$.
At sufficiently high temperatures one expects that $g(T)$ will be very
small, therefore the electric mass will be much smaller than the
temperature scale and the magnetic mass even smaller. This is not the
case however in realistic lattice simulations, since at
high temperatures the finite-size-effects become big, thus restricting
the investigation to relatively low temperatures.

There have been two recent determinations of the electric and magnetic
masses; the first one \cite{karsch1} measured correlators of gauge
invariant objects in QCD, yielding $m_{mag} \approx (2.9 \pm 0.2)~T,~~m_{el}
\approx (1.4 \pm 0.2)~T$. The second one \cite{karsch2} is based on
the gluon propagators themselves in $SU(2)$ gauge theory and yields
$m_{mag} = 0.466(15)~g^2(T)~T,~~m_{el} = 2.484(52)~T$.

On the other hand, there has been a
different interpretation \cite{gyulassy} of the results of
\cite{karsch1}, claiming
that the gluon propagators do not really behave as decreasing exponentials,
but have instead a power law behaviour at small distances, which would mean
that the results of \cite{karsch1} are compatible even with a vanishing value
for the magnetic mass.

An approach that could shed some light on the space-time dependence of
the gluon propagator is the study of momentum space propagators
(\cite{bernard,mart}). This has
the advantage that one studies the whole momentum space propagator, rather 
than the few numbers which survive the sum over the hyperplanes.
Moreover it opens the possibility to study the behaviour in
each momentum region separately and relate the momentum under study to the
scales that enter the problem, which is not possible in configuration
space studies. Finally, as shown in \cite{bernard}, there is the very important
technical advantage of a much better behaved covariance matrix. In
particular, the
covariance matrix for configuration space propagators is singular for the
lattice sizes presently used; this is not the case in momentum space,
offering the possibility to perform fully correlated chi-squared fits.

\section {The finite temperature gluon propagator}
\noindent

At finite temperature any second rank tensor may be expanded 
in the basis of the following four tensors:
\be
P_{\mu\nu}^T = \delta_{\mu i} (\delta_{i j} -{ {k_i k_j} \over 
{{\bf k}^2} }) \delta_{j \nu}
\ee
\be
P_{ \mu \nu} ^L =(\delta_{\mu 4}- { {k_4 k_\mu} \over {k^2} } )
                 { {k^2} \over {{\bf k}^2}}
                 (\delta_{\nu 4}- { {k_4 k_\nu} \over {k^2} } )
\ee
\be
P_{ \mu \nu} ^G = { {k_\mu k_\nu} \over {k^2} }
\ee
\be
P_{ \mu \nu} ^S = { 1 \over \sqrt{2 {\bf k}^2} }
                  ( k_\mu (\delta_{\nu 4}-{ {k_4 k_\nu} \over {k^2} })
                 +( k_\nu (\delta_{\mu 4}-{ {k_4 k_\mu} \over {k^2} }) ) 
\ee
In the {\em Landau gauge} the free gluon propagator is of the form:
$D_{\mu \nu}(k)={ {-1} \over {k_4^2+{\bf k}^2} }(P_{\mu \nu}^L +P_{\mu \nu}^T).$
If one expands the self energy tensor $ \Pi_{\mu \nu}$ on the above basis:
\be
\Pi_{\mu \nu}^{a b} (k_4,{\bf k})={\bf \Pi}_L^{a b} P^L_{\mu \nu}
+{\bf \Pi}_T^{a b} P^T_{\mu \nu}
+{\bf \Pi}_S^{a b} P^S_{\mu \nu}
+{\bf \Pi}_G^{a b} P^G_{\mu \nu}.
\ee
Taking into account that ${\bf \Pi}_J^{a b}({\bf k},k_4) = \delta^{a
b} \Pi_J({\bf k},k_4)$, where the index $J$ may be $L$, $T$,
$S$ or $G$,
one may show that the full
gluon propagator at finite temperature is given by the expression:
\beba
G_{\mu \nu}^{a b} ({\bf k},k_4) & \equiv <A_\mu^a A_\nu^b> \\
 & =\delta^{a b} ({ {1} \over {k_4^2+{\bf k}^2+ \Pi_L ({\bf k},k_4)} }
P^L_{\mu \nu} +
{ {1} \over {k_4^2+{\bf k}^2+ \Pi_T ({\bf k},k_4) } } P^T_{\mu \nu})
\eaee
Now we observe that $P_{4 \mu}^T(k)=P_{\mu 4}^T (k) = 0,~ \mu=1,...4 $ 
for any  value 
of k, while $P^L_{i j}(k)=k_4^2 { {k_i k_j} \over {k^2~{\bf k}^2 }},
~ i,j=1,2,3$,
which vanishes for all i, j when $k_4 = 0$. This fact, that the static
spatial self-energy is always transverse, may be shown \cite{gross}
using the Ward identities
satisfied by $G_{\mu \nu}^{a b}$ making no appeal to a specific gauge.
In the sequel we stick to the static case, $k_4=0$.
The limits of $\Pi_L({\bf k},k_4=0)$ and
$\Pi_T({\bf k}, k_4=0)$ as ${\bf k}$ goes to zero are the
electric and magnetic masses (squared) respectively.
In this work we are going to measure the (lattice versions of the)
quantities
\be
C_{j}({\bf k}) \equiv \frac{1}{2} \sum_{a,b=1}^2 \delta_{a b} G^{a b}_{j j}
({\bf k}, k_4=0) \equiv G ({\bf k}, k_4=0) P^T_{j j} ({\bf k},
k_4=0),
\ee
where j runs from 1 to 3 and no sum over j is implied. In fact, we
do not fit $C_{j}({\bf k})$, but the quantities
\be
G
({\bf k}, k_4=0) = \frac{1}{ {\bf k}^2+ \Pi_{T} ({\bf k}, k_4=0)}.
\ee

The infrared non-perturbative behaviour of the gluon propagator has
been studied using a variety of methods, with conflicting conclusions.
Other works \cite{ball} predict for the propagator a very singular
behaviour at zero {\bf k}, like $\frac{1}{({\bf k}^{2})^{2}}$ and a
confining property is conjectured; on the other side, there are claims
\cite{zwanziger} that the tree-level pole at ${\bf k}^2=0$ is removed,
due to non-perturbative effects and the Green's function vanishes at
zero momentum. It is a fact that in the infrared region there should
appear a dynamically generated mass $M(g, \mu)$. 
However, the
generation of this mass cannot be seen in perturbation theory, since
it must behave like $M(g,\mu) \approx \mu e^{-\frac{constant}{g^2}}$
for very small $g$, that is it should exhibit an essential singularity
at zero gauge coupling. It is expected that {\em
non-perturbative} effects may yield negative powers of the momentum in
the vacuum polarization function of the gluon, introducing various
mass scales:
\be
\Pi ({\bf k}^2) = { {m^2(g,\mu, T)} \over {{\bf k}^2} } +{
{b^4(g,\mu, T)} \over {({\bf k}^2)^2} } + \dots
\ee
g is the coupling constant and $m(g,\mu, T)$, $b(g,\mu, T)$ have
dimensions of mass and depend non-analytically on g. We have only
shown the first two terms of an expansion in $\frac{1}{{\bf k}^2}$,
since they correspond to well-known suggestions about the gluon
propagators. In particular, if $b(g,\mu, T)=0$, a non-zero $m(g, \mu,
T)$ gives rise to a mass pole in the gluon propagator. On the other
hand, if $m(g, \mu, T)=0$, a non-zero $b(g,\mu, T)$ will give rise to
a propagator of the form
\be
G(k)={ { {\bf k}^2} \over {({\bf k}^2)^2+b^4} },
\ee
which has been proposed by Gribov \cite{gribov}.

\section {The method}
\noindent

We measured correlations of the gauge potential, defined through the relation:
\be
A_\mu (x) \equiv { {U_\mu(n)-U_\mu^\dagger(n)} \over {2i} },~\mu=1,2,3,4.
\ee
The various expectation values should be calculated in a lattice version of
the Landau gauge, implemented by making the quantity
\be
\Sigma[\Lambda] \equiv \sum_{n,\mu} Tr[\Lambda(n) U_\mu(n)
\Lambda^\dagger (n+\mu)] \label{sig}
\ee
a global maximum with respect to the gauge transformations $\Lambda(n)$.
This guarantees not only the satisfaction of the lattice version of
the condition $\partial_\mu A^\mu =0$, but also the additional constraint
imposed
by Gribov. Let us note that the lattice Faddeev-Popov operator is just
the Hessian matrix of $\Sigma[\Lambda]$ with respect to $\Lambda[n]$. The
maximization condition fixes the sign of the Hessian to be the correct
one. Moreover, the maximization condition also enforces the smoothness 
of the continuum limit. Of course, one cannot say whether this
proposal really gets rid of all of the Gribov copies; however, it is 
easily seen that the Gribov copies of the trivial configuration are
eliminated, since the latter is the only constant configuration which 
maximizes $\Sigma[\Lambda]$. Thus, it may be expected that also the
problem in its general form may be less severe. 

In order to calculate the gauge dependent correlators, we updated with
the usual Wilson action and transformed the resulting configurations to the 
Landau gauge before taking measurements. This can be proved
\cite{mandula} to incorporate
the effects of the Faddeev-Popov determinant. For SU(2), which we are
considering, we sweep through the lattice and at each lattice site
we calculate analytically the gauge transformation that maximizes the sum
of the links beginning or ending at this site. Of course, this gauge
transformation will disturb the gauge condition on the neighboring
sites, so we expect that the algorithm relaxes to the global maximum we
are looking for after several sweeps.
A good gauge fixing is very important for the reliability of the
results. A straightforward check is to calculate the $<A_3(x) A_3(0)>$
correlator in configuration space, summed over the directions 1, 2 and
4: this quantity should be constant, as a consequence of the gauge
fixing. We have considered the gauge fixing as good enough if the
variation of this correlator was not larger than 0.1 percent. This in
turn dictated that the quantity $\frac{1}{N^3 N_t}
\sum_{k,n,\mu}(\partial_\mu A^\mu_k(n))^2$ (which must vanish in
the Landau gauge) should be less than about $10^{-5}$. This last
requirement has been practically used as the criterion to stop the
gauge fixing iterations. Let us note that the number of measurements
done varied between 3000 and 5000 for each point; we also mention
that a number of overrelaxation sweeps (~5) was performed between 
successive Monte Carlo steps.
In figure 1 we show the configuration space correlators $<A_1(x)
A_1(0)>$ and $<A_3(x) A_3(0)>$ versus $x$. The first correlator decays
as usual, while the second one is constant, because of the gauge
fixing, as just explained.

\section {Results and conclusions}
\noindent

Let us first explain the way of analyzing our results. We have
measured the correlators of link variables $A_j~~(j=1,2,3)$ 
on the lattice and
then performed the Fourier transform, considering only the case
$k_4=0$. The lattice versions of the propagators have been considered
in the fit function $G ({\bf k}^2) = \frac{C}{ \hat {\bf k}^2 +
\mu^2}$ ($\mu \equiv m \alpha$), which means that we used everywhere 
the dimensionless lattice
version $\hat {\bf k}$ of the three-momentum rather than the usual one
${\bf k}$. Let us recall the relevant definitions: $\hat p_i a \equiv
\hat k_i \equiv 2 sin \frac {k_i a} {2}$ where $k_i a = \frac{2
\pi}{N} n_i, n_i=-\frac{N}{2} + 1, \dots, \frac{N}{2}$ for the even N
we have been using. 
We have considered lattices with two different
temporal extents ($N_t=2$ and $N_t=3$) to get some feeling about the
finite size effects due to the smallest of the lattice dimensions. Of
course the corresponding momentum $k_4$ has been set to zero, as an
external momentum, however it also appears as an internal momentum in
the self-energy graphs and may very well influence the results. In
most simulations the spatial dimension $N$ of the lattice has been
$10$, however we have also used $N=12$ and $N=14$ in some cases.

We have chosen to measure the momentum $\hat p$ ($\hat p \equiv | \hat
{\bf p} |$, $\hat k \equiv |\hat {\bf k}|$) in units of the
temperature $T$ of each lattice, since this is one of the dominant
scales of the problem (along with $g(T) T$ and $g^2(T) T$). We note
that $\frac{\hat p_{max}}{T} = N_t \hat k_{max}$, so the range of the
quantity $\frac{\hat p}{T}$ explored by each lattice is proportional
to its ``temporal'' extent, provided $N$ is kept constant. Taking this fact
into account, the (lattice) momentum range $[0,\hat k_{max}]$ 
has been divided into $N_t$
parts and the gluon propagators have been considered over the intervals: $
\left[ \hat k_{max} \frac{n-1}{N_t},\hat k_{max} \frac{n}{N_t} \right],
n=1, \dots, N_t $, for the quantity $\hat k$, corresponding to the
intervals $\left[ (n-1) \hat k_{max}, n \hat k_{max} \right], n=1, \dots,
N_t $ for the quantity $\frac{\hat p}{T}$. We note that this division of
the intervals is arbitrary; one might divide the intervals in a different
number of subintervals (bigger or smaller than $N_t$). We observe that a
specific value of n yields the same values of $\frac{\hat p}{T}$ no matter
which lattice we are considering. In particular, the value n=1 corresponds
to the results we are going to depict in the figures and has to do with the
interval $ 0 \leq \frac{\hat p}{T} \leq \hat k_{max}$.  Note that for the
lattices $10^3 \times 2$, at most ten momenta may be considered in the
fits, while for the lattices $10^3 \times 3$ the corresponding number of
momenta is four. In particular for the $N_t = 2$ case we observed that the
qualitative results we refer to in the following do not change if we
consider shorter intervals within the ones corresponding to the various
values of n; there are minor changes in the values of the mass and one gets
smaller values for the chis squared. There is a serious exception, though,
for $n=1$: the point $\hat {\bf k}={\bf 0}$ should be included in the fit in
any case, since the second smallest momentum possible ($\frac{2 \pi}{N
\alpha}$) is already of the order of $T$.

We have measured the propagators for several values of
$\frac{T}{T_c}$, ranging from $0.60$ to $6.0$ (the values greater than
$4.0$ could only be reached on the lattices with $N_t=2$; for $N_t=3$
the required $\beta_g$ would be too large, given the spatial dimensions
of the lattices we have been using).

In figures 2a, 2b, we show the results for the spacelike propagators
(magnetic sector) as a function of the temperature for $N_t = 2,~3$
respectively, at ``low" momenta (n=1). We plot the output for
$\frac{T}{m} = \frac{1}{N_t \mu}$ versus $\frac{T}{T_c}$, 
as well as curves fitting these data . We have also
included in the figures the data for $\frac{T}{T_c} < 1$. These have
nothing to do with the magnetic mass, of course, however it is worth
noting that no dramatic effects occur in the values of these screening
masses at the phase transition point. One may observe that lattices
with $N_t=2$ and $N_t=3$ give values quite close for equal
$\frac{T}{T_c}$ (with some small discrepancies 
around $\frac{T}{T_c} = 2$).

To construct the fitting functions we invoke the expected behaviour
$m=c g^2(T) T$ of the magnetic mass along with  the ansatz
$g^{-2}(T)=\frac{11}{12 \pi^2} log(\frac{T}{\lambda T_c})$, inspired
by (the leading order in) the renormalization group equation for the
gauge coupling. In practice we have fitted $\frac{T}{m}$ to the form
$A~ log(\frac{T}{T_c}) + B$ and determined $c, \lambda$ through
$c=\frac{11}{12 \pi^2 A}, \lambda=e^{-\frac{B}{A}}$. The fit has taken
into account only the data with $\frac{T}{T_c} > 1.6$. It turns out
that $$ \begin{array}{ll} c(N_t=2)=0.20(3),& \lambda(N_t=2)=0.55(15) \\
c(N_t=3)=0.26(2),& \lambda(N_t=3)=0.26(8).\end{array} $$ These values for $c$
are more or less in agreement to the early publications \cite{bls} on
the subject, where $c$ has been found $0.24$.

Let us note that using the above expression for $g(T)$ and the just
mentioned values of $\lambda$, we may find the ratio of the magnetic 
mass to the temperature:

$$ m(N_t=2)=1.9 T,~~~m(N_t=3)=1.6 T$$

For even larger momenta ($n=2,~n=3$) the mass is zero in all cases.
Thus, at high momenta the correlators tend to free field ones.

It is of interest to check what happens to the quantities measured
above in the case we have a three-dimensional lattice, which would
correspond to the ``infinite temperature" limit. In figure 3 we show
the results for the resulting masses for a three - dimensional
($12^3$) lattice. On dimensional grounds one expects that any mass
should be proportional to $g_3^2$, the square of the 
three-dimensional gauge
coupling. We find that $$m=0.49(3) g_3^2.$$ If we make the (crude)
identification of the {\em (bare)} $g_3^2$ to $g^2(T) T$, 
we see that the three-dimensional
result would imply a much larger value for the quantity $c$ than the
one found from the four-dimensional lattices; this result is, in fact,
close to the result given in \cite{karsch2}.

Let us note that, if we also treat $2 \times 2$ {\em plaquette} correlators
in the same way as the link correlators, we get $m_{pl}=2.3(5) g^2_3$
(in agreement with \cite{fks}), that is a much larger factor than
the one extracted from the {\em link} variables.

{\bf Acknowledgements:} The author wants to thank the DESY theory
group for its warm hospitality during the early stages of this work.
In particular, I thank I.Montvay for giving me his simulation code for
$SU(2)$ gauge fields, R.Sommer for providing an efficient random
generator and W.Buchmueller, Z.Fodor, M.Reuter and K.Farakos for
useful discussions. A very enlightening discussion with F.Karsch is
gratefully acknowledged.

\newpage

\begin{center}
{\bf Figure captions} 
\end{center}

\vspace{0.5cm}

{\bf Figure 1:} The correlators $<A_1(x) A_1(0)>$ (diamonds) and
$<A_3(x) A_3(0)>$ (crosses) versus $x$ for $\frac{T}{T_{c}}=2.5$,
$N=10$ and $N_t=2$.

{\bf Figure 2:} $\frac{T}{m}$ versus $\frac{T}{T_{c}}$ for the
space-like link correlators with n=1 in the cases $N_t=2$ (figure 2a)
and $N_t=3$ (figure 2b). The fitting curves are also plotted.

{\bf Figure 3:} $m \alpha$ versus $G \equiv g_3^2 \alpha$ 
for the link correlators
with n=1 in the three-dimensional case. The fitting curve is also
plotted.

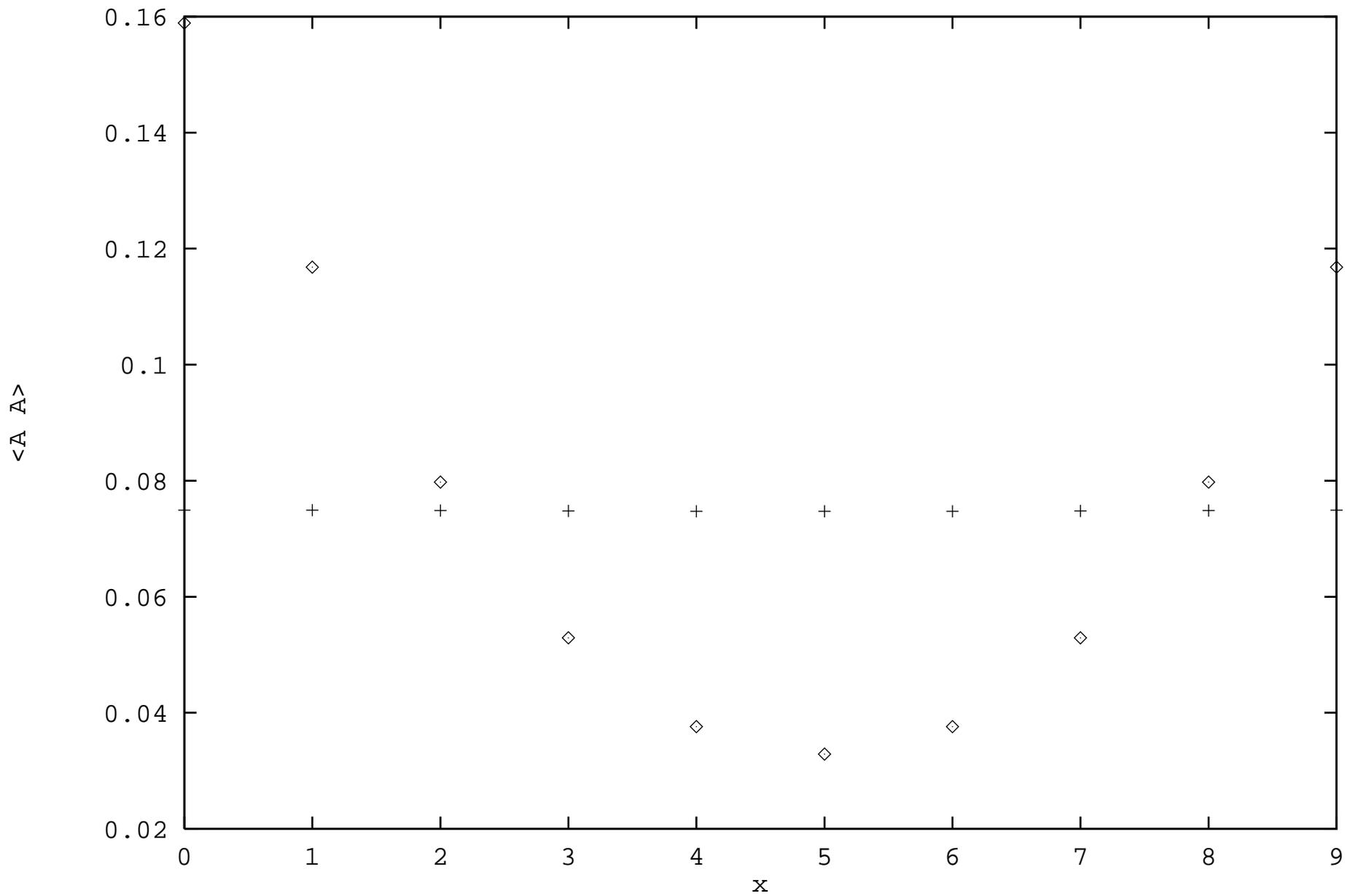
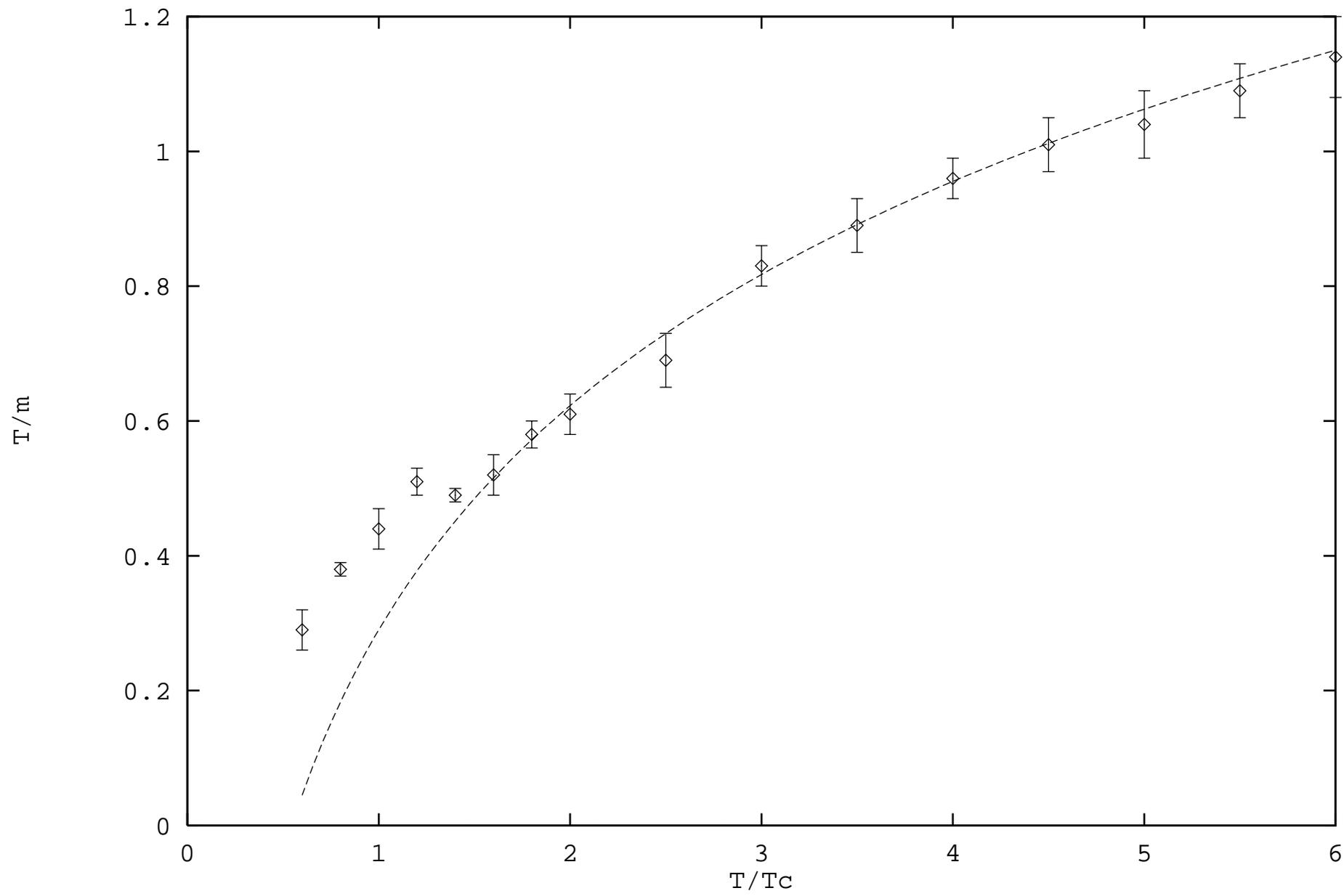
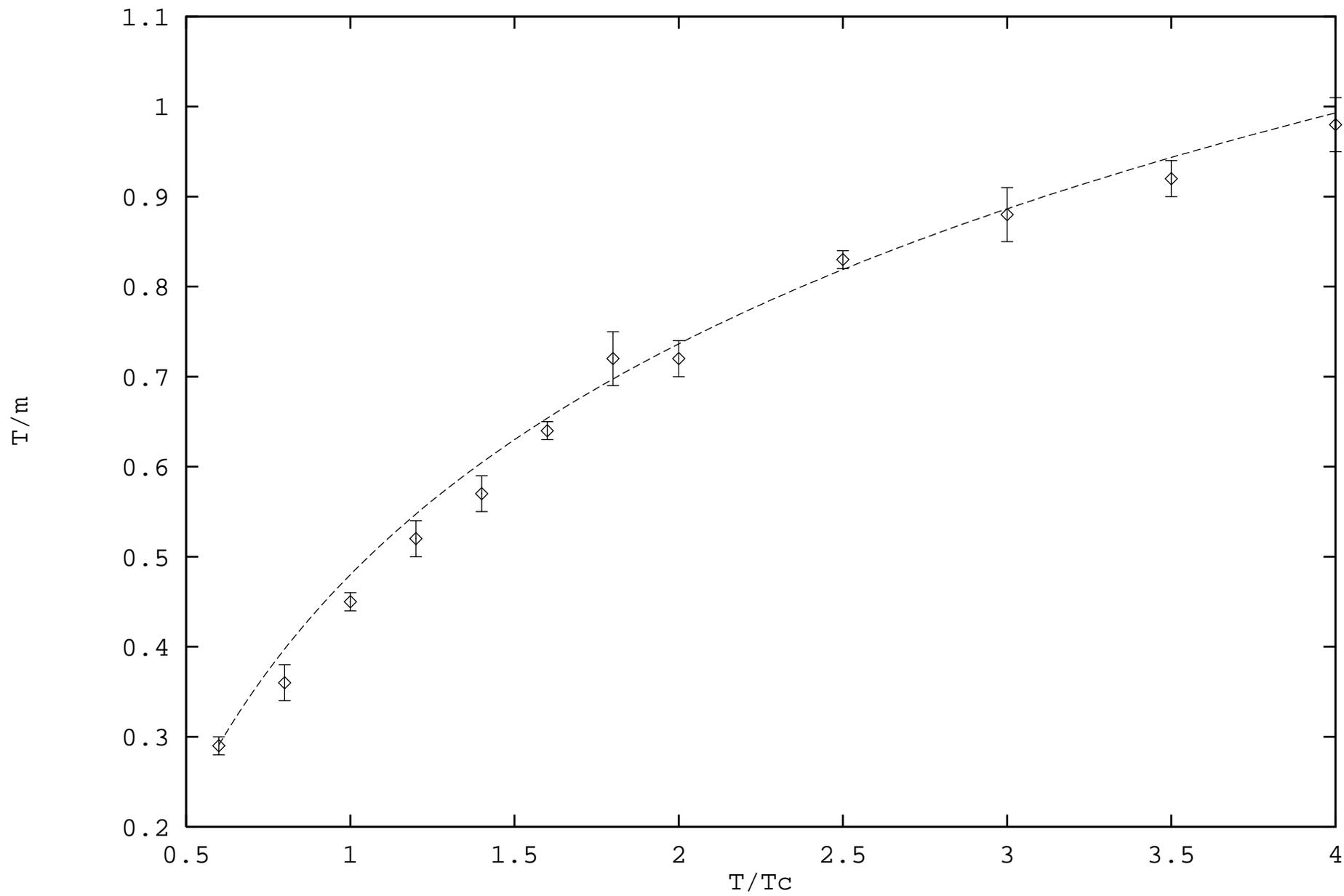
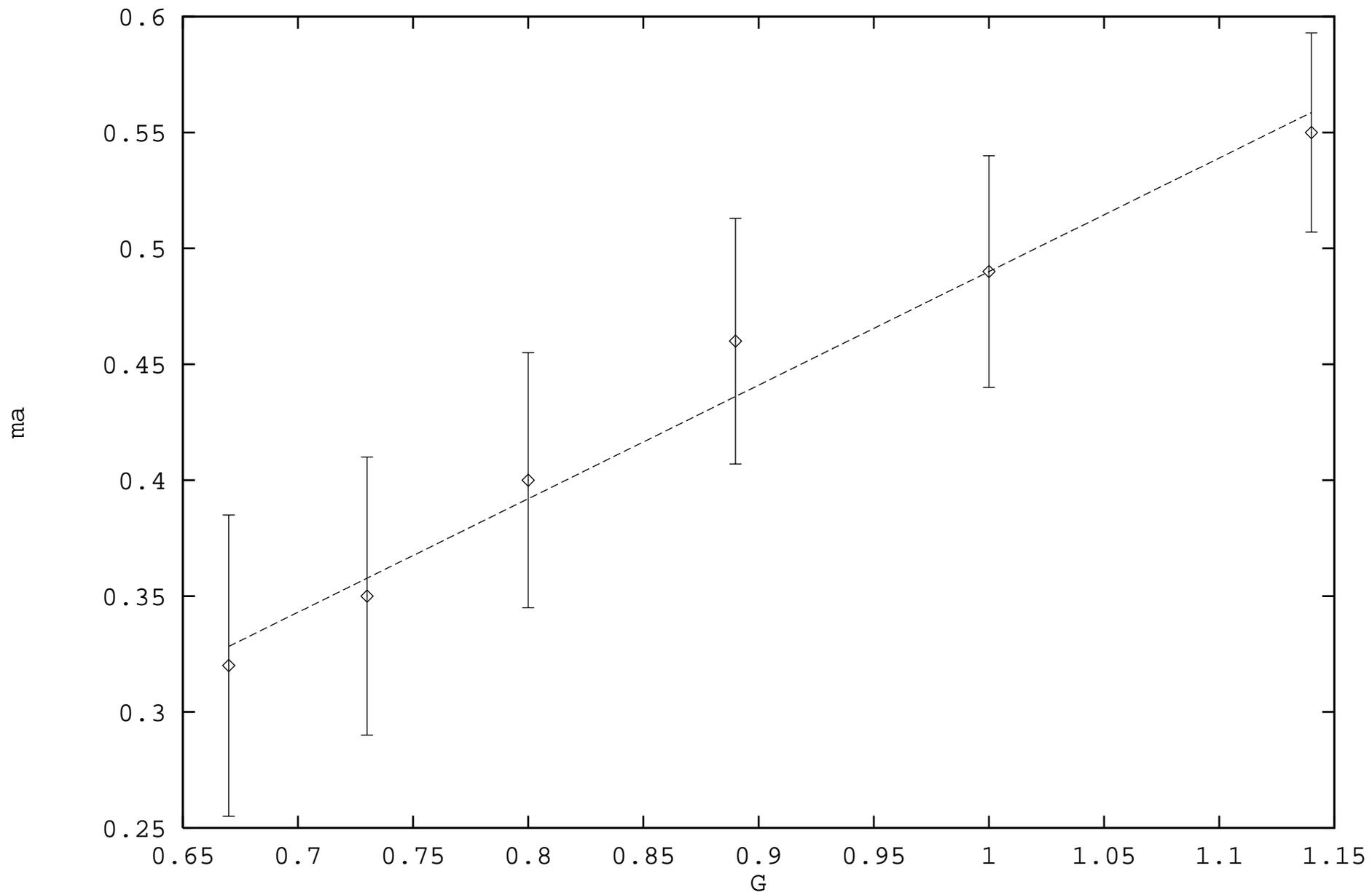
\end{document}